\documentclass{article}

\usepackage{arxiv}

\usepackage[utf8]{inputenc} %
\usepackage[T1]{fontenc}    %
\usepackage{hyperref}       %
\usepackage{url}            %
\usepackage{booktabs,longtable}       %
\usepackage{siunitx}
\usepackage{amsfonts}       %
\usepackage{nicefrac}       %
\usepackage{microtype}      %
\usepackage{lipsum}
\usepackage{xcolor}
\usepackage{float}

\providecommand{\tightlist}{%
	\setlength{\itemsep}{0pt}\setlength{\parskip}{0pt}}
\usepackage{mathtools}
\usepackage{float}
\usepackage[style=nature,backend=biber,doi=false,url=false]{biblatex}
\title{d-SEAMS: Deferred Structural Elucidation Analysis for Molecular
	Simulations}

\author{
	\href{https://orcid.org/0000-0002-2393-8056}{\includegraphics[scale=0.06]{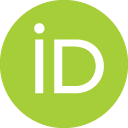}\hspace{1mm}Rohit
    Goswami}\thanks{Currently at the Department of Chemistry, IIT Kanpur,
    \textit{equal contribution}} \\
	Department of Chemical Engineering\\
	Indian Institute of Technology Kanpur\\
	\texttt{rgoswami@iitk.ac.in} \\
	\And
	\href{https://orcid.org/0000-0001-8706-2383}{\includegraphics[scale=0.06]{orcid.png}\hspace{1mm}Amrita Goswami}\thanks{\textit{equal contribution}} \\
	Department of Chemical Engineering\\
	Indian Institute of Technology Kanpur\\
	\texttt{amritag@iitk.ac.in} \\
	\And
	\href{https://orcid.org/0000-0001-8056-2115}{\includegraphics[scale=0.06]{orcid.png}\hspace{1mm}Jayant K. Singh}\thanks{\textbf{Corresponding Author}} \\
	Department of Chemical Engineering\\
	Indian Institute of Technology Kanpur\\
	\texttt{jayantks@iitk.ac.in} \\
}

\begin{document}
\maketitle

\begin{abstract}
	Structural analyses are an integral part of computational research on
  nucleation and supercooled water, whose accuracy and efficiency can
  impact the validity and feasibility of such studies. The underlying
  molecular mechanisms of these often elusive and computationally
  expensive processes can be inferred from the evolution of ice-like
  structures, determined using appropriate structural analysis techniques.
  We present d-SEAMS, a free and open-source post-processing engine for
  the analysis of molecular dynamics trajectories, which is specifically
  able to qualitatively classify ice structures, in both strong
  confinement and bulk systems. For the first time, recent algorithms for
  confined ice structure determination have been implemented, along with
  topological network criteria for bulk ice structure determination.
  Recognizing the need for customization in structural analysis, d-SEAMS
  has a unique code architecture, built with \texttt{nix}, employing a
  \texttt{YAML}-\texttt{Lua} scripting pipeline. The software has been
  designed to be user-friendly and easy to extend. The engine outputs are
  compatible with popular graphics software suites, allowing for immediate
  visual insights into the systems studied. We demonstrate the features of
  d-SEAMS by using it to analyze nucleation in the bulk regime and for
  quasi-one and quasi-two-dimensional systems. Structural time evolution
  and quantitative metrics are determined for heterogenous ice nucleation
  on a silver-exposed \(\beta\)-AgI surface, homogenous ice nucleation,
  flat monolayer square ice formation and freezing of an ice nanotube.
\end{abstract}


\hypertarget{introduction}{%
\section{Introduction}\label{introduction}}

The increase in power and efficiency of high-performance computing
resources have enabled researchers to directly observe nucleation events
in molecular-dynamics simulations. Concomitantly, the determination of
ice-like structures from simulations is essential for the interpretation
of nucleation events, since trajectories provide positional data of
particles, and do not explicitly track crystal structures and defects
\cite{stukowskiStructureIdentificationMethods2012}. Nucleation of
soft-matter is further complicated by the emergence of competing ice
polymorphs \cite{Carignano2007, Herlach2016} with small free energy
differences \cite{Woodcock1997}. The ice-like structures formed are also
continually distorted by thermal fluctuations, which locally disrupt
long-range order. These issues can make automated and accurate
structural determination intractable, especially for weakly crystalline
regions \cite{Reinhart2017}. Surface interactions and confinement can
strongly influence nucleation behaviour
\cite{Lupi2014, Cox2015, Bi2017}, which add to the complexity of the
structure determination problem.

Water is a deceptively simple molecule, exhibiting rich and complex
phase behaviour in bulk and confinement
\cite{petrenko2002physics, Salzmann2006,
Salzmann2006a, Salzmann2009, Falenty2014, Rosso2016, Algara-Siller2015,
Agrawal2016, Zhao2014, Takaiwa2007, Bai2010}. At least \(17\) bulk ice
polymorphs have been observed experimentally
\cite{Salzmann2011, Salzmann2019}. Water confined within nanometer
length scales exhibits even more diversity, forming ordered
hydrogen-bond networks of ice nanotubes, monolayers, bilayers and
trilayers \cite{Chen2016, Koga2000, Zangi2003, Bai2012, Zhao2014,
zhuCompressionLimitTwoDimensional2015, Zhu2016, gaoPhaseDiagramWater2018}.
The structural determination of the several possible polymorphs of
water, in bulk and in confinement, is of crucial importance in the
qualitative and quantitative analysis of simulation data. The analysis
of the evolution of ordered structures during a nucleation event is
desirable, since it can reveal details of the underlying molecular
mechanism and provide important physical insights into the system.

In this work, we present d-SEAMS, a cohesive post-processing structural
analysis engine, which is capable of coherently classifying water
structures under strong confinement and in bulk systems, from quasi-one
and two-dimensional confinement to bulk ice polymorphs. Confined ice
polymorphs are often identified ``by eye'' wherein the hydrogen-bond
network (HBN) is manually inspected \cite{Bai2010, Zhao2014,
zhuCompressionLimitTwoDimensional2015}. In particular, d-SEAMS automates
the process of structural analysis and qualitative metric calculation,
eliminating the need for visual inspection.

Conflicting package clashes due to transitive or indirect dependencies
is a recurring problem in software development and use
\cite{Boettiger2015}. Resolving such issues can often be non-trivial,
relying on removal of unnecessary dependencies, system updates (for most
Linux clusters) and other manual strategies that may be ineffective for
a complex dependency tree. These manual strategies are increasingly
difficult to carry out on restricted access machines with high up-times,
which are common to the HPC (High Performance Computing) clusters, used
for such computationally demanding simulations.

d-SEAMS circumvents the problem of `dependency hell' by using
\texttt{nix} \cite{dolstraNixSafePolicyFree2004} for generating
reproducible dependency build-graphs \cite{prins2008nix}. Users can run
d-SEAMS using the exact build-environment of the developers and
vice-versa, bypassing installation and use issues on various systems and
HPC clusters, ensuring reproducible results.

The parameters of structural analysis techniques are often tweaked to
suit the specific requirements of the desired study
\cite{Wolde1996,liHomogeneousIceNucleation2011}. d-SEAMS has been
designed to permit easy extensions to the code and the implementation of
custom work-flows. The engine simultaneously incorporates user-friendly
interface functions without compromising on functionality. Key features
implemented include a primitive ring analysis algorithm
\cite{Franzblau1991}, topological network criteria for bulk
\cite{haji-akbariDirectCalculationIce2015} and confined systems
\cite{Goswami2019} and new qualitative order parameters. Popular
analysis techniques, including bond-orientational parameters
\cite{Steinhardt1983, Wolde1996} and related criteria
\cite{mooreFreezingMeltingStructure2010, nguyenIdentificationClathrateHydrates2015},
have also been implemented. Outputs are produced in formats compatible
with popular visualization softwares, including OVITO
\cite{Stukowski2009} and VMD \cite{Humphrey1996}. We describe how
d-SEAMS has been used to analyze and characterize ice nucleation on an
ice-promoting AgI surface, homogenous bulk nucleation, the formation of
monolayer ice and the freezing of a quasi-one-dimensional ice nanotube
(INT).

\hypertarget{code-architecture}{%
\section{Code Architecture}\label{code-architecture}}

\hypertarget{nix-expressions-for-reproducible-builds}{%
\subsection{Nix Expressions for Reproducible
Builds}\label{nix-expressions-for-reproducible-builds}}

Computational software implemented in declarative package management
systems suffer from design concerns for the end-users and the
developers. The developers have the onus to package their software as
per the many imperative systems (Ubuntu and \texttt{apt}, RedHat and
\texttt{yum}, ArchLinux and \texttt{pacman}). Additionally, the users
must ensure that the interrelated dependencies match perfectly. In
essence, the issue is that the packages and versions at the time of
build are not guaranteed automatically at the user's end, even if the
software is packaged appropriately for the operating system. This issue
stems from the fact that the configuration after installation is the
result of a series of stateful transformations which cannot be
reproduced \cite{dolstraNixOSPurelyFunctional2010}. We have opted to
package our software as a nix-derivation
\cite{dolstraNixSafePolicyFree2004}, which also provides a functionally
reproducible environment for reproducible bug-tests. Since the build
system is essentially a static graph of build actions, the environment
produced is a boon for reproducibility, avoiding circular build-time
dependencies and incomplete dependency specifications
\cite{pengReproducibleResearchComputational2011}. More details on the
design rationale are in the Electronic Supplementary Information.

\hypertarget{pipeline-of-work-flows}{%
\subsection{Pipeline of Work-flows}\label{pipeline-of-work-flows}}

\begin{figure}
\centering
\includegraphics[scale=0.5]{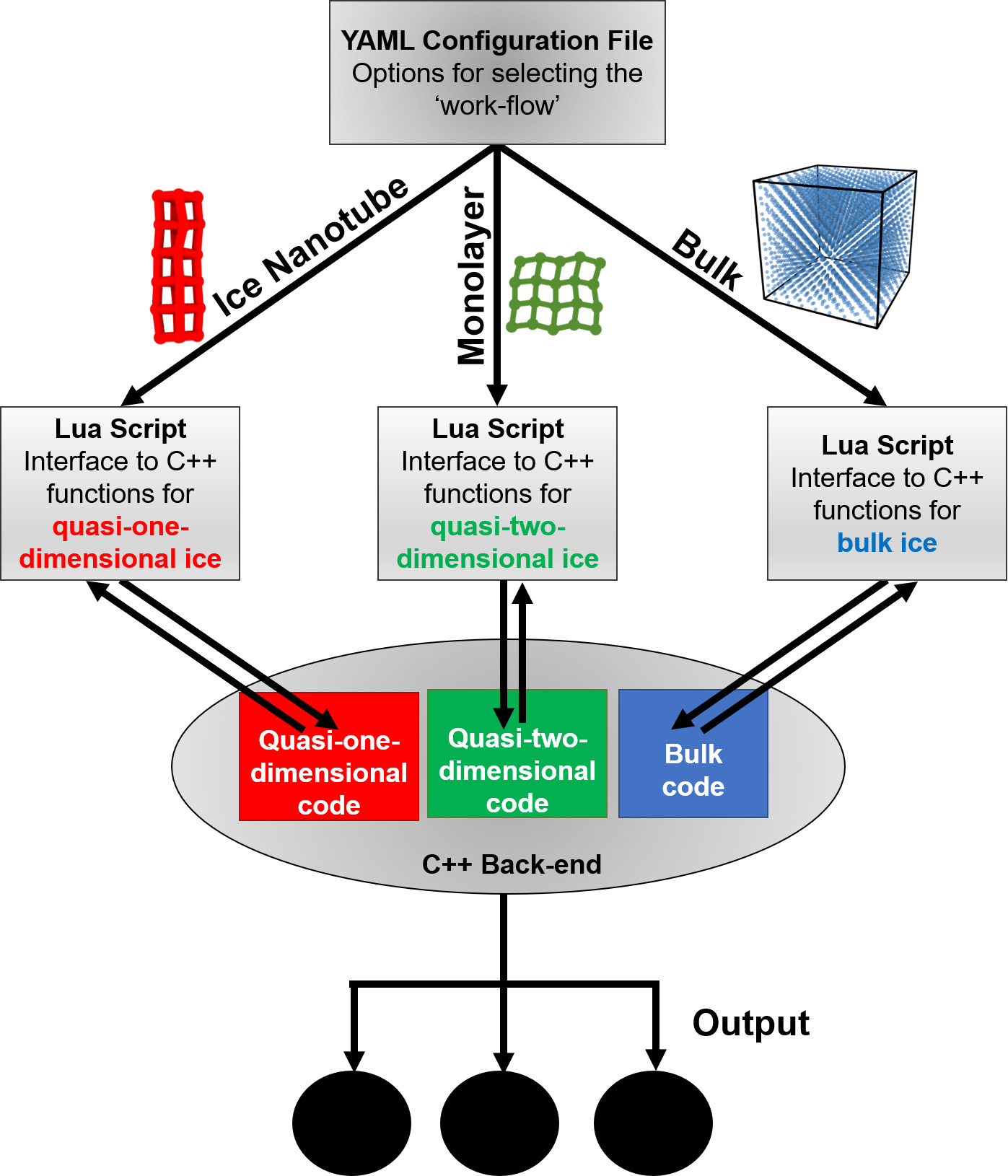}
\caption{Code pipeline of d-SEAMS. The `YAML` script file provides options to choose the type of the system and the corresponding work-flow, for either bulk, quasi-one-dimensional quasi-tqo-dimensional monolayer sytems. The `Lua` interface functions interact with the `C++` functions for each work-flow, and functions of other work-flows are not exposed to the user. \label{fig:workflow}}
\end{figure}

Figure \ref{fig:workflow} visually depicts the logical pipeline of
typical work-flows in d-SEAMS. An input system can either be a bulk
system, a quasi-one-dimensional system or a quasi-two-dimensional
system. The criteria and analysis algorithms tend to differ across
scales as well. Thus, we have separated the mutually exclusive
work-flows for bulk, quasi-one-dimensional and quasi-two-dimensional
systems, with separate modular code blocks for each type of system. The
general software pipeline is organized as follows:

\begin{itemize}
\tightlist
\item
  The user chooses whether the system is a bulk system,
  quasi-one-dimensional or quasi-two-dimensional system, by setting the
  appropriate values in the \texttt{YAML} input file.
\item
  The \texttt{Lua} script calls the functions to read in, analyze and
  output the desired results.
\item
  Simulation data is read from LAMMPS trajectory files. The XYZ file
  format is also currently supported. Other file formats may be
  converted into an ASCII LAMMPS trajectory format \cite{Shirts2016}. 
\item
  Options for controlling the output results can be specified in the
  \texttt{Lua} file input.
\item
  Output directories are created automatically, according to the chosen
  work-flow.
\end{itemize}

\hypertarget{structural-identification-features}{%
\subsection{Structural Identification
Features}\label{structural-identification-features}}

We are able to apply arbitary transformations on the particle
collections, in an efficient and idiomatic manner. Currently we have
implemented the following structural schemes based on this easily
extensible framework:

\begin{itemize}
\tightlist
\item
  The bond orientational parameters \cite{Steinhardt1983, Wolde1996},
  CHILL \cite{mooreFreezingMeltingStructure2010} and CHILL+
  \cite{nguyenIdentificationClathrateHydrates2015} parameters have been
  implemented. Clustering of ice-like molecules \cite{Stoddard1978},
  based on these parameters can be optionally carried out by the user.
  The largest ice cluster so obtained can also be re-centered, for ease
  of visualization.
\item
  Topological network criteria for bulk ice determination
  \cite{haji-akbariDirectCalculationIce2015} have been implemented.
  d-SEAMS is able to identify and write out detailed information about
  Double-diamond cages (DDCs), hexagonal cages (HCs), and mixed rings
  for every frame. Additionally, the number of basal and prismatic rings
  are also computed.
\item
  Primitive rings are identified. First, all possible rings are found
  using an exhaustive backtracking algorithm, following which
  non-shortest path rings are removed \cite{Franzblau1991}. Ring
  networks with only hydrogen-bonded connections can be optionally
  determined.
\item
  Confined quasi-two-dimensional ice classification by topological and
  graph theoretic approaches to the hydrogen-bonded ice-like particles.
\item
  Quasi-one dimensional ice nano-tube (INT) and quasi-two-dimensional
  monolayer ice classification via topological network criteria
  \cite{Goswami2019}. The building blocks of \(n\)-gonal prismatic ice
  are explicitly and unequivocally identified.
\item
  d-SEAMS is capable of calculating geometric order parameters for
  describing the phase transitions in confined ice.
\end{itemize}

\hypertarget{applications}{%
\section{Applications}\label{applications}}

Using d-SEAMS, structural and qualitative information as a time series
can be easily obtained. This makes it possible to study a variety of
diverse nucleating systems. Even the mechanism of nucleation growth can
be inferred from an ensemble of trajectories, by analyzing the
qualitative metrics supported by d-SEAMS.

\hypertarget{bulk-systems}{%
\subsection{Bulk Systems}\label{bulk-systems}}

\hypertarget{heterogenous-nucleation-on-an-ice-promoting-surface}{%
\subsubsection{Heterogenous Nucleation on an Ice-Promoting
Surface}\label{heterogenous-nucleation-on-an-ice-promoting-surface}}

Silver iodide is an effective ice nucleating agent, whose lattice
closely resembles that of bulk ice \cite{Vonnegut1947}. The smooth
Ag-exposed \(\beta\)-AgI surface has been known to promote
layer-by-layer growth of hexagonal ice (Ih) and cubic ice (Ic) in
simulations \cite{Zielke2014}. Using d-SEAMS, we probe the underlying
mechanism of the ice nucleation growth and behaviour. Ten independent
simulations of \(5120\) TIP4P/Ice \cite{Abascal2005} molecules on a
free-standing AgI surface were run upto \(\approx 200 \ ns\) at
\(240 \ K\). The simulation setup is similar to that of previous work in
the literature \cite{prernaStudyIceNucleation2019}.

Here we employ a topological network criterion
\cite{haji-akbariDirectCalculationIce2015} for identifying the building
blocks of Ih and Ic, called hexagonal cages (HCs) and double-diamond
cages (DDCs), respectively. Rings which belong to both HCs and DDCs are
classified as mixed rings.

\begin{figure}
\centering
\includegraphics[width=\textwidth]{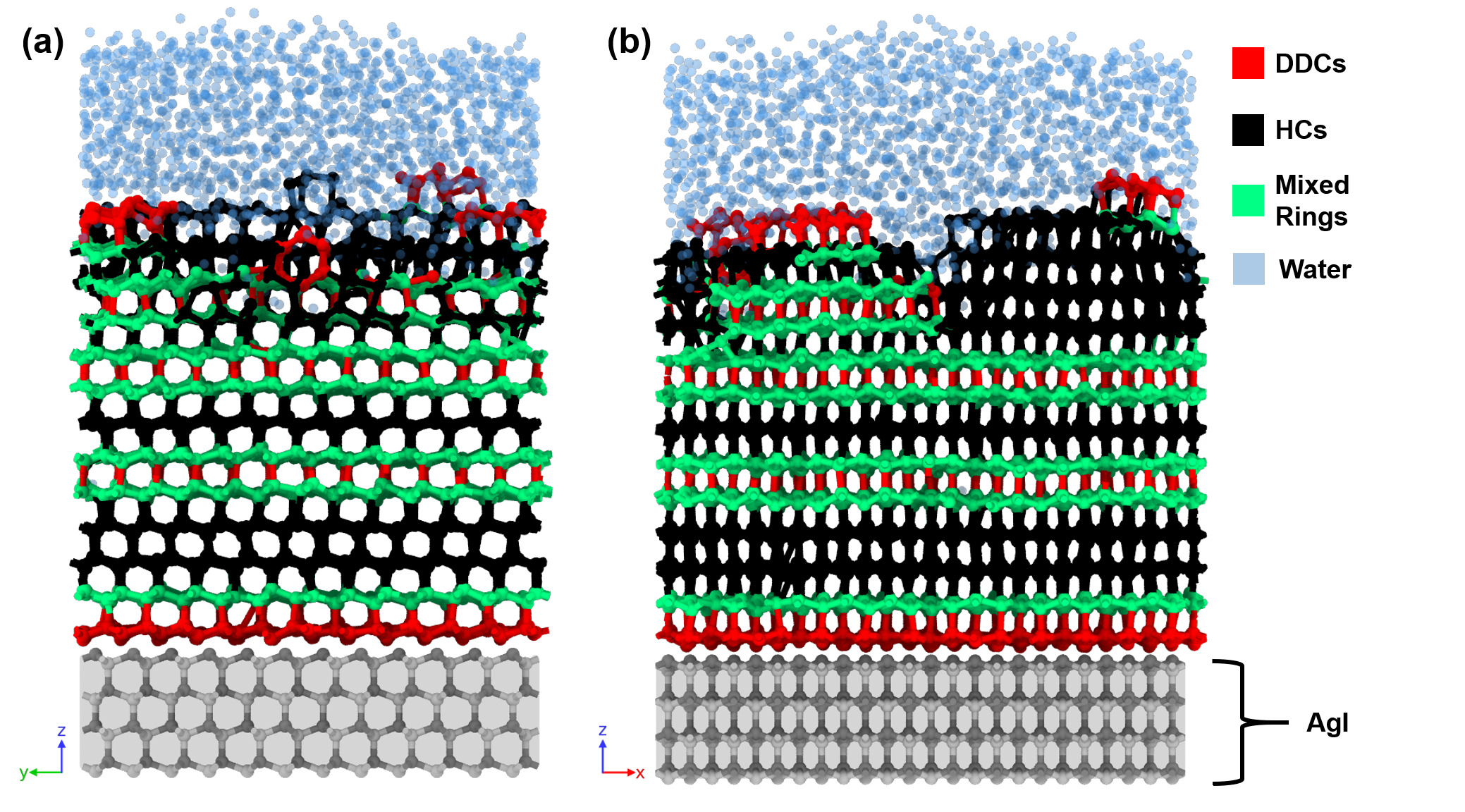}
\caption{(a) $YZ$-plane and (b) $XZ$-plane views of layered hexagonal ice
(Ih) and cubic ice (Ic) after $200 \ ns$. DDCs, HCs and mixed rings are coloured
in red, black and green, respectively. The silver and iodine atoms are dark-grey
and grey respectively, with the AgI sheet highlighted in light grey.
\label{fig:fullAgIsys}}
\end{figure}

Figure \ref{fig:fullAgIsys} depicts layers of HCs and DDCs, growing from
the AgI surface, after \(200 \ ns\) of simulation time. The close
lattice match between the AgI surface and the ice phases is clearly
visible. However, the CHILL algorithm
\cite{mooreFreezingMeltingStructure2010} identifies the first layer of
molecules close to the surface as liquid. It was noted in previous
studies that although the CHILL algorithm does not identify the first
layer of water molecules as ice, these molecules actually form part of
the initial ice layer \cite{Zielke2014, Zielke2016}. The topological
network criterion used here correctly classifies these water molecules
as constituents of HCs or DDCs.

\begin{figure}
\centering
\includegraphics[width=0.6\textwidth]{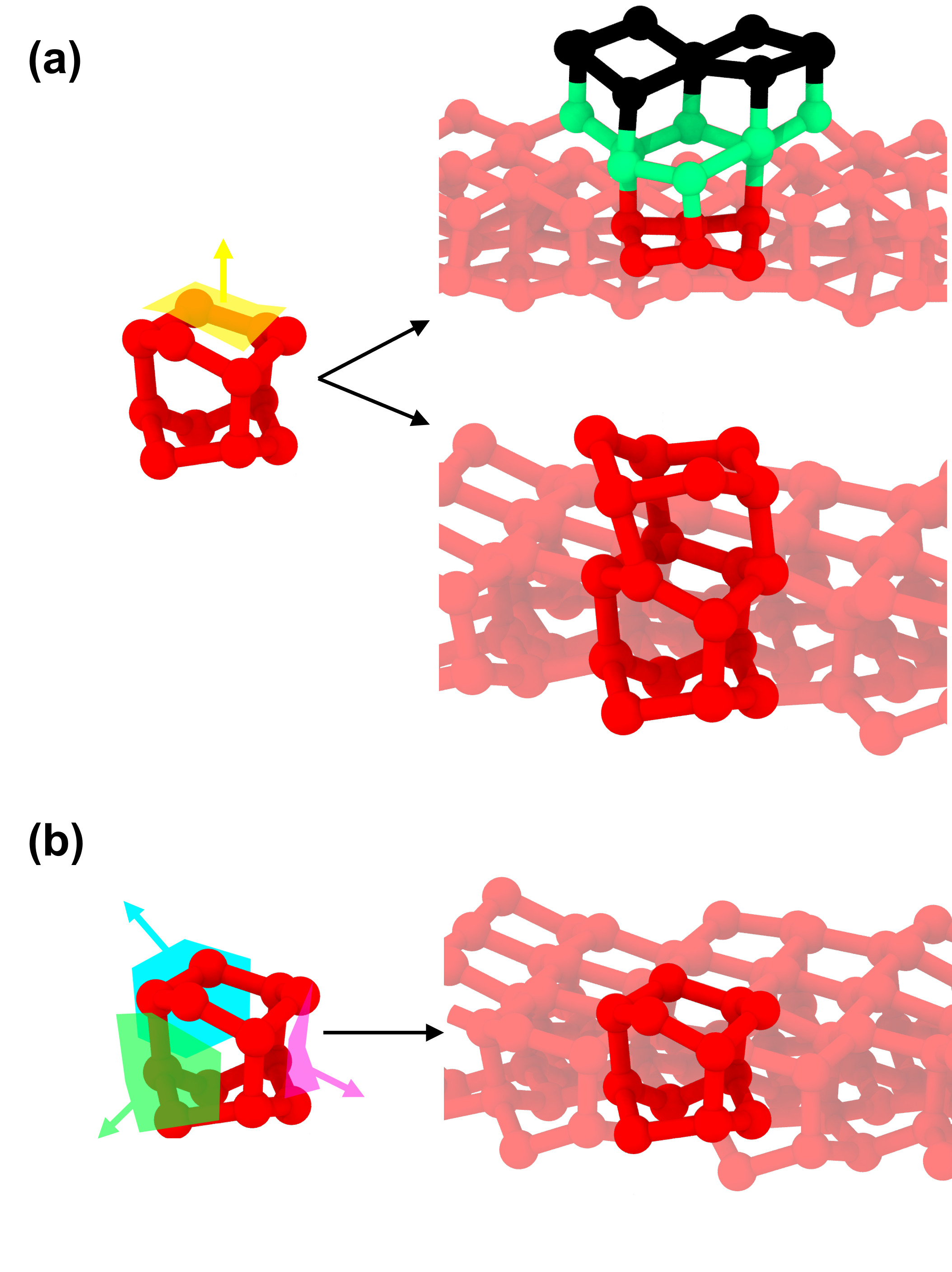}
\caption{(a) Attachment of cages on the upper basal face of an HC. The
basal face has been highlighted in yellow. Basal faces of an HC can support the
growth of both DDCs (black) and HCs (red). (b) Attachment of HCs on the three
prismatic faces, shaded in blue, green and mauve, of an HC. The prismatic faces
of an HC cannot support the growth of DDCs. The colour scheme is the same as in
Figure \ref{fig:fullAgIsys}. \label{fig:planeGrowth}}
\end{figure}

The Ag-exposed \(\beta\)-AgI surface supports the attachment of the
basal plane of HCs, as well as DDCs. Figure \ref{fig:planeGrowth} shows
how layers grow through attachment of HCs and DDCs.

The peripheral rings of DDCs can support the attachment of both HCs and
DDCs. However, the prismatic planes of an HC (highlighted in blue, green
and mauve in Figure \ref{fig:planeGrowth}(b)) can only support the
anchoring of HCs. Since the HCs grow upwards from the AgI surface,
through the basal plane (highlighted in yellow in Figure
\ref{fig:planeGrowth}(a)), only HCs can grow from the prismatic planes
of pre-existing HCs in each layer. We surmise that the layer-wise growth
of HCs in the lateral dimensions is because of this growth behaviour.
This is in keeping with the stacking effects observed in the literature
\cite{prernaStudyIceNucleation2019}.

\begin{figure}
\centering
\includegraphics[width=0.8\textwidth]{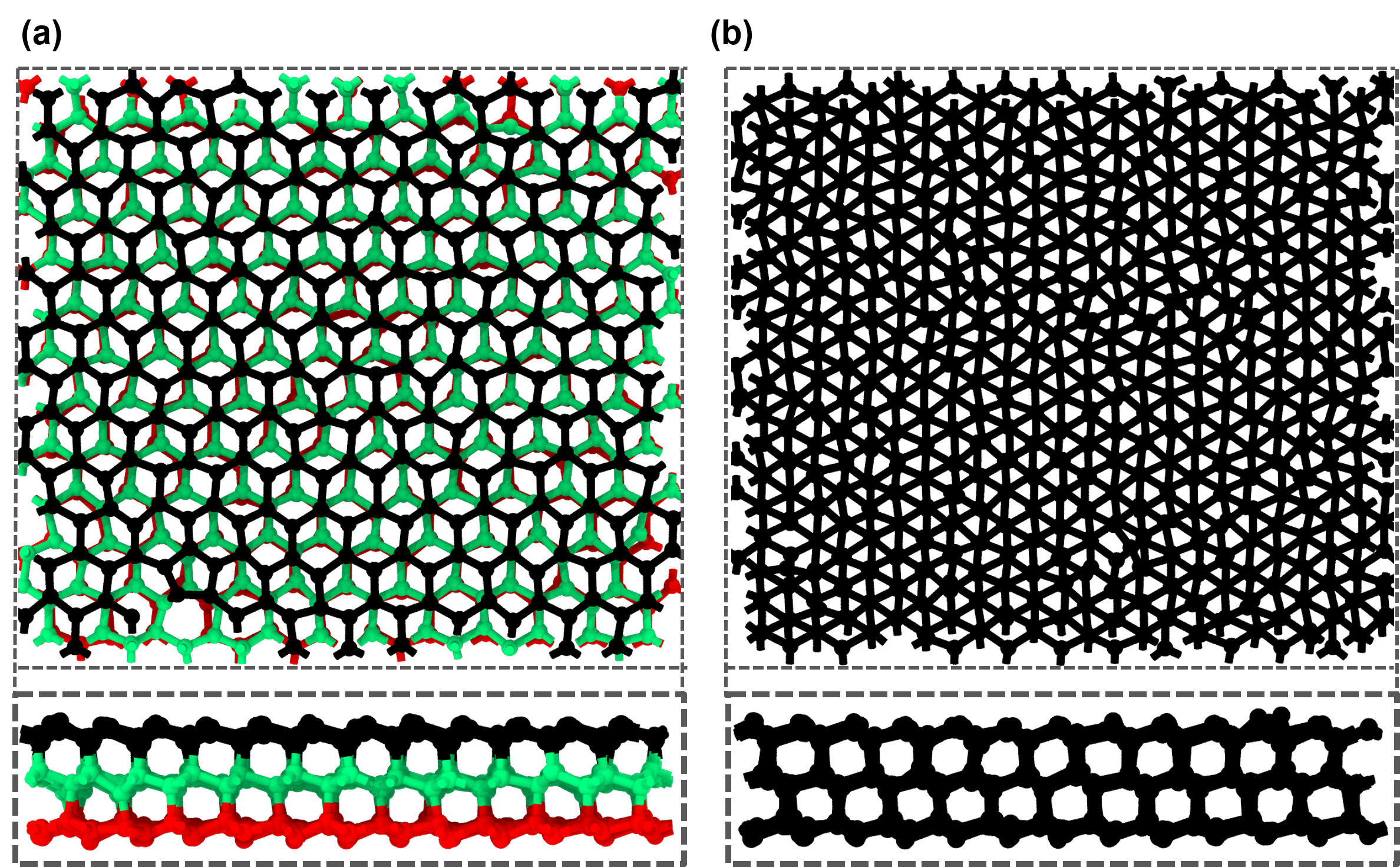}
\caption{Overhead and side views showing the stacking arrangements of the
first two layers of cages on the AgI surface, when more than half the water has
been converted to ice, after more than $\approx 150 \ ns$ of simulation time.
(a) The first layer is a layer of HCs. The complete second layer is a layer of
DDCs in all cases observed. Mixed rings are sandwiched between the first layer
of HCs and the second layer of DDCs. (b) The first and second layers are DDCs.
The colour scheme is the same as in the preceding figures.
\label{fig:stackViews}}
\end{figure}

The effect of the surface is the most significant within the first two
layers of cages. In all the independent trajectories, a single layer of
either HCs or DDCs is formed first, upon which a second layer of DDCs
grows. The two possible outcomes of the stacking of cages, observed in
independent simulations, are depicted visually in Figure
\ref{fig:stackViews}. Figure \ref{fig:stackViews}(a) shows the overhead
view of the stacking arrangement when the first layer and second layer
are HCs and DDCs, respectively. Figure \ref{fig:stackViews}(b) shows the
other stacking arrangement observed, wherein the first and second layers
are exclusively composed of DDCs. Both types of arrangements seem
equally likely, based on the roughly equal proportions of stacking
observed in all the independent simulations performed. Regardless of the
composition of the first two layers, alternating layers of varying
widths of HCs and DDCs are formed in all the simulations. Interestingly,
although HCs sometimes grow on the first layer, by the end of the
simulations, the second layer is exclusively composed of DDCs.

We track the growth of the first two layers, with time, for a particular
trajectory, whose stacking at the end of \(200 \ ns\) is shown in Figure
\ref{fig:stackViews}(a).

\begin{figure}
\centering
\includegraphics[scale = 0.45]{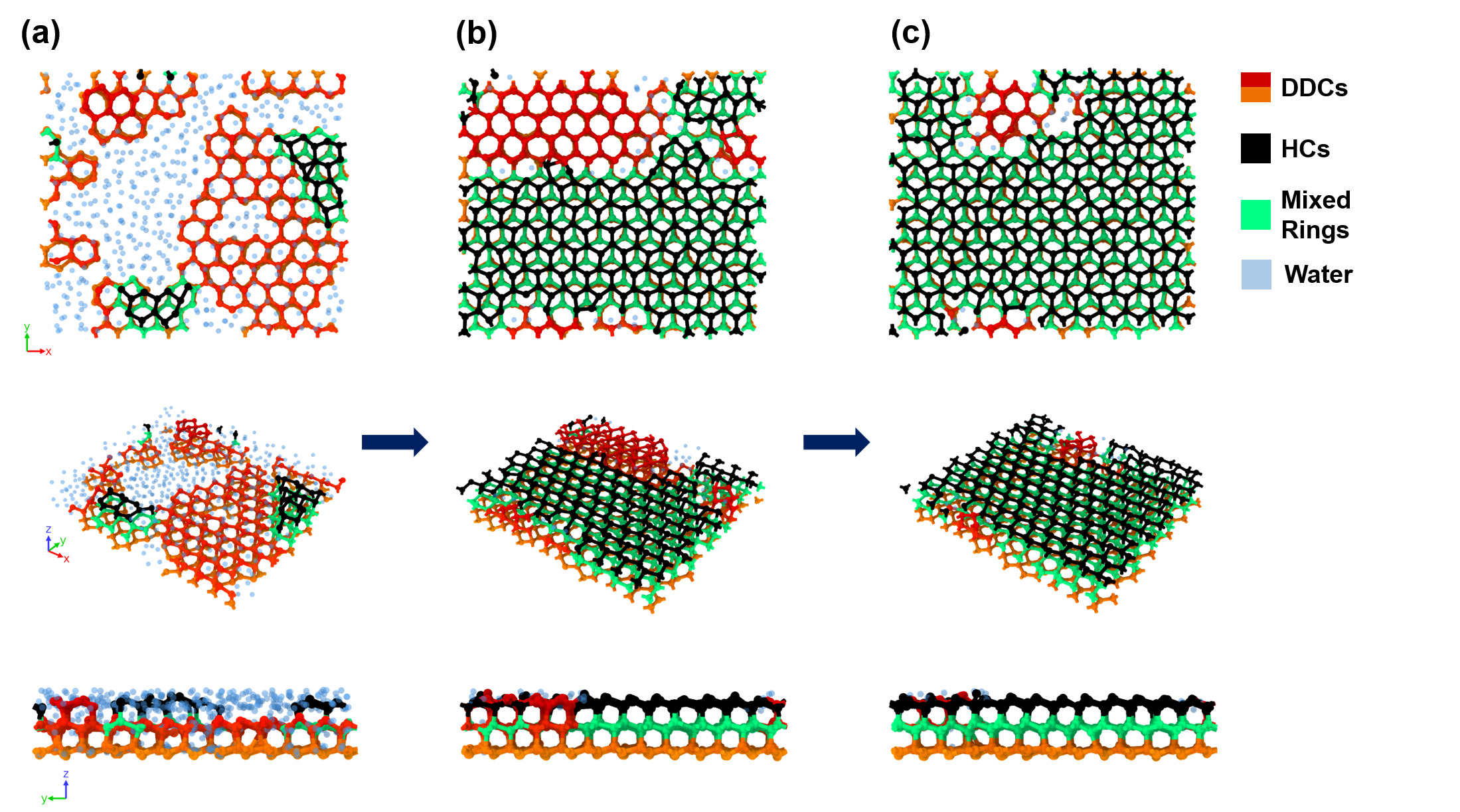}
\caption{Formation of layers of HCs and DDCs within the first $100 \ ns$
of simulation time. (a) After $\approx 14 \ ns$, the first layer has not
completely formed. (b) After $\approx 60 \ ns$, part of the second layer is
comprised of HCs, and a separate portion of the second layer is composed of
DDCs. The part of the second layer made up of HCs is distinguishable in the top
view because of the AA-stacking order of HCs on the first layer of HCs. (c)
After $\approx 100 \ ns$, the second layer is almost entirely composed of DDCs.
The colour scheme is shown in the legend. DDCs are coloured in shades of yellow
to vermilion, according to the distance from the AgI surface.
\label{fig:evolSheet}}
\end{figure}

Figure \ref{fig:evolSheet} shows the growth of the first two layers of
cages within the first \(100 \ ns\) of simulation time. After
\(\approx 60 \ ns\), the second layer is partially composed of HCs and
DDCs (Figure \ref{fig:evolSheet}(b)). The HC portion is not anchored to
the DDC part of the second layer, since DDCs cannot grow on the exposed
lateral prismatic faces of the HCs. We note that eventually, by the time
that the second layer of cages becomes continuous, HCs in the second
layer disappear and are completely replaced by DDCs. In fact, the
eventual layering of HCs and DDCs in the first two layers from the AgI
surface is also observed on void-defect incorporated Ag-exposed
\(\beta\)-AgI surfaces \cite{prernaStudyIceNucleation2019}.

\hypertarget{homogenous-nucleation-growth-of-the-largest-ice-cluster}{%
\subsubsection{Homogenous Nucleation: Growth of the Largest Ice
Cluster}\label{homogenous-nucleation-growth-of-the-largest-ice-cluster}}

\begin{figure}
\centering
\includegraphics[width=0.8\textwidth]{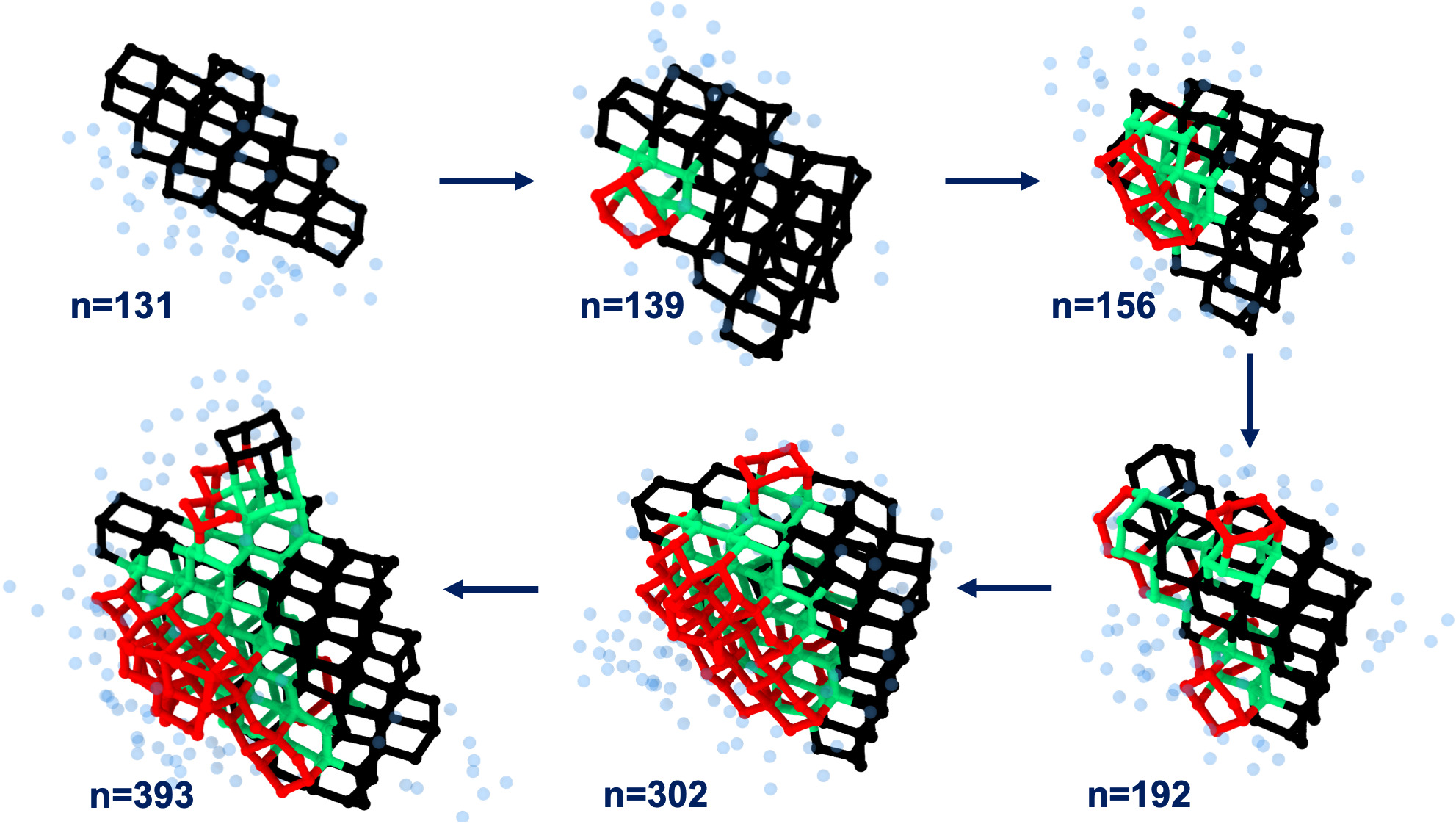}
\caption{Evolution of an ice crystallite of mW water at $208 \ GPa$ in
the NPT ensemble. DDCs are in black, HCs are in red and mixed rings belonging to
both DDCs and HCs are in green. The number of particles in the largest cluster
$n$ is shown for every snapshot of the trajectory. Water molecules which do not
participate in the cages are in transparent blue. \label{fig:iceEvol}}
\end{figure}

Here, we have analyzed a trajectory which exhibits a successful
homogenous nucleation event, in the bulk phase. Independent simulations
of \(4096\) particles, modelled using the monoatomic water (mW) water
model \cite{Molinero2009}, were equilibriated at \(300 \ K\) and
quenched to \(208 \ K\). Simulations were subsequently run at constant
temperature and pressure, in the NPT ensemble, at \(1 \ atm\) and
\(208 \ K\) for upto \(1 \ \mu s\). Figure \ref{fig:iceEvol} shows the
structural features and growth of the largest ice cluster. In our
ensemble of trajectories, we have observed that the largest ice
crystallites which survive and grow to the post-critical size are rich
in DDCs (shown in black in Figure \ref{fig:iceEvol}). We find that even
in post-critical crystallites, the core of the crystallites remains rich
in DDCs. These findings match those gleaned from extensive forward-flux
sampling simulations of both mW water and TIP4P/Ice
\cite{haji-akbariDirectCalculationIce2015}. The growth of the largest
crystallite with time, showing fluctuations in the cluster size with
time, is depicted in a supplementary movie, entitled
\texttt{largestClusterGrowth.avi}, in the ESI (electronic supplementary
information).

\hypertarget{quasi-two-dimensional-systems}{%
\subsection{Quasi-Two-Dimensional
Systems}\label{quasi-two-dimensional-systems}}

\begin{figure}
\centering
\includegraphics[width=1.0\textwidth]{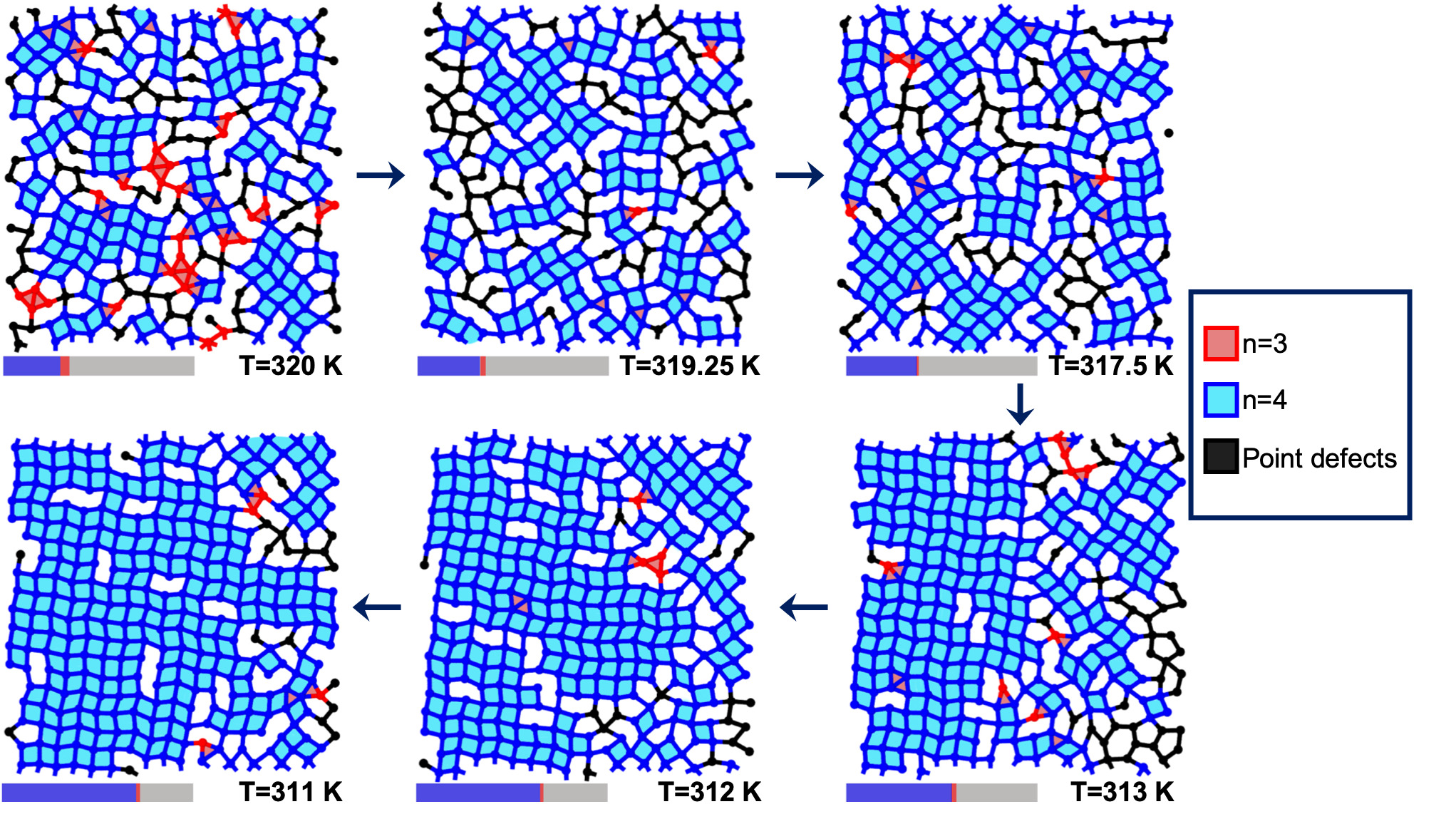}
\caption{Snapshots of monolayer water being cooled from $320 \ K$ to $310
\ K$, at $1 \ GPa$. The stacked bars in the lower left corner of each snapshot
denote the coverage area percentages for the 4-membered rings (blue) and
3-membered rings (red). The colours of the particles, bonds and interior ring
shading are shown in the legend.\label{fig:monoSnap}}
\end{figure}

We further analyze Flat Monolayer Square Ice (fMSI), using d-SEAMS. FMSI
has been observed in simulations, when water subjected to a lateral
pressure of \(1 \ GPa\) is sandwiched between rigid graphene sheets,
separated by a slit width of \(6\) \si{\angstrom}
\cite{zhuCompressionLimitTwoDimensional2015, Yang2017}. d-SEAMS is
capable of calculating the coverage area metric \cite{Goswami2019}, an
area-based order parameter, at every timestep of a trajectory. We
demonstrate the efficacy of the coverage area metric in describing phase
transitions, during the cooling of a quasi-two-dimensional monolayer
from \(320 \
K\) to \(310 \ K\) over \(1 \ ns\). The visuals produced were generated
from the output of d-SEAMS. Figure \ref{fig:monoSnap} shows snapshots of
the monolayer water through various stages of cooling, exhibiting a
gradual increase in the proportion of the 4-membered rings. At
\(320 \ K\), the coverage area of the 4-membered rings is
\(\approx 70 \%\), which indicates that the monolayer is predominantly
fMSI. This phase change is depicted in a supplementary video
(\texttt{monolayer.avi}) in the ESI.

\begin{figure}
\centering
\includegraphics[width=0.7\textwidth]{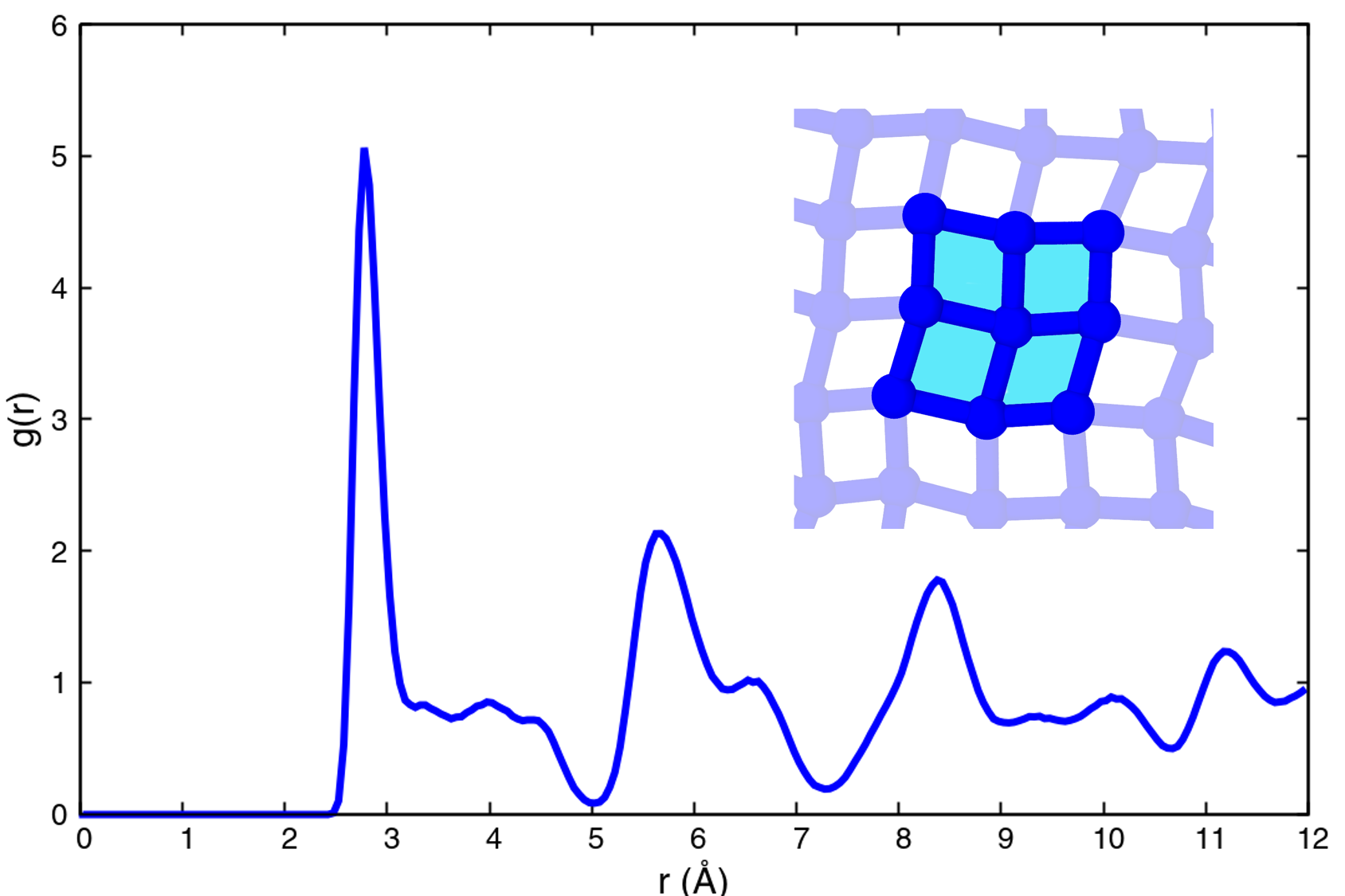}
\caption{In-plane oxygen-oxygen radial distribution function of the ice
formed at $310 \ K$ and $P_{xx} = 1 \ GPa$. The inset shows the 4-membered rings
characteristic of fMSI. \label{fig:rdfFMSI}}
\end{figure}

d-SEAMS is also capable of calculating the in-plane radial distribution
function. The solid-like ordered nature of the ice formed at \(310 \ K\)
is typified by the shape of the 2-dimensional oxygen-oxygen radial
distribution function in Figure \ref{fig:rdfFMSI}. The \(g_{OO}(r)\) has
the highest peak at \(2.775 \pm 0.05\) \si{\angstrom}, followed by a
shoulder peak and a third smaller peak; the same features have been
previously observed for fMSI. The frames were obtained from an
equilibriated \(1 \ ns\) run at a constant temperature of \(310 \ K\)
and lateral pressure of \(1 \ GPa\).

\begin{longtable}[]{@{}ccc@{}}
\caption{Ring statistics and Coverage Area percentage for triangular and
square ices \label{ringStat}}\tabularnewline
\toprule
Number of nodes & \(Coverage \ Area_n \ \%\) & Number of
rings\tabularnewline
\midrule
\endfirsthead
\toprule
Number of nodes & \(Coverage \ Area_n \ \%\) & Number of
rings\tabularnewline
\midrule
\endhead
n=4 & \(74.422 \pm 3.345\) & \(238.94 \pm 10.723\)\tabularnewline
n=3 & \(0.877 \pm 0.5\) & \(6.2 \pm 3.544\)\tabularnewline
\bottomrule
\end{longtable}

Figure \ref{ringStat} summarizes the ring statistics and coverage area
metric values, determined by d-SEAMS, for the fMSI structure formed at
\(310 \ K\). The coverage area metric is a more intuitive and more
stable metric than the number of rings, as is evident from the data. The
number of rings is more sensitive to small thermal fluctuations which
are not relevant to the overall classification scheme.

\hypertarget{quasi-one-dimensional-systems}{%
\subsection{Quasi-One-Dimensional
Systems}\label{quasi-one-dimensional-systems}}

\begin{figure}
\centering
\includegraphics[width=0.8\textwidth]{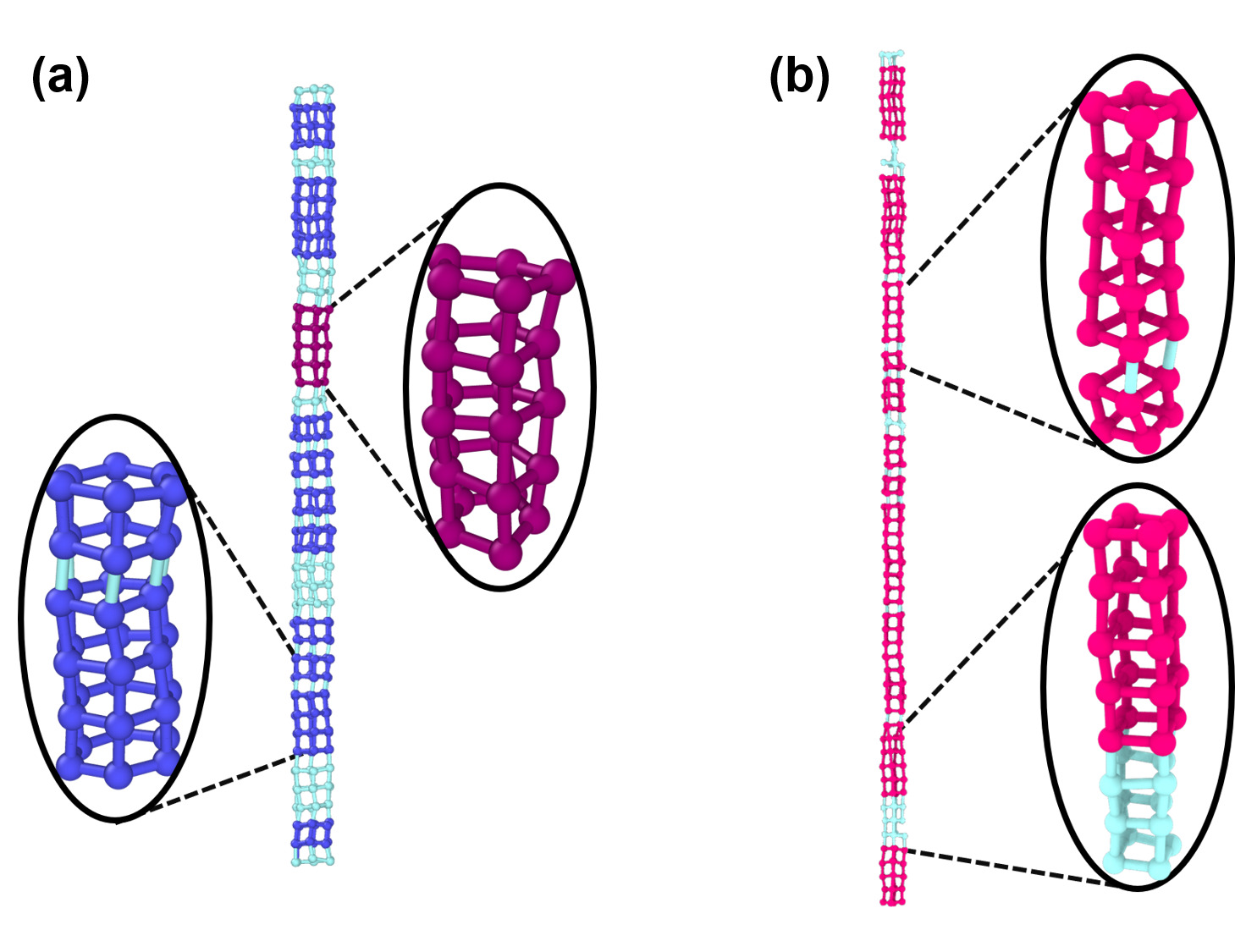}
\caption{View of ice structures at $240$K confined within (a) ($13,0$)
zig-zag nanotube at $0.1$MPa and (b) ($11,0$) zig-zag nanotube at $1$MPa. The
unclassified phase, either water or deformed prism blocks, is sea-green. Prism
blocks of pentagonal, hexagonal, and tetragonal ices are shown in purple, blue
and pink respectively.\label{fig:someLab}}
\end{figure}

The perspective views of ice nanotubes (INTs) are shown in Figure
\ref{fig:someLab}, for water constrained within a \((13,0)\) and
\((11,0)\) zig-zag smooth featureless single-walled nanotube (SWNT). For
the SWNTs, \(R=13\) and \(R=11\) correspond to diameters of \(10.1\)
\si{\angstrom} and \(8.6\) \si{\angstrom}, respectively.

At an axial pressure of \(P_{zz}=0.1 MPa\), primarily hexagonal prism
blocks are formed within the INT in Figure \ref{fig:someLab}(a). A small
proportion of the ice nanoribbon is comprised of pentagonal prism
blocks, while the rest are deformed prism blocks classified as an
`unclassified' phase. The unclassified phase may denote deformed prisms,
the liquid phase or hollow helices, which are not explicitly
differentiated using the prism classification scheme. In this particular
case, the hollow nature of the nanoribbon and the interspersed positions
of the unclassified phase between prism blocks, indicate that the
unclassified phase is comprised of deformed prisms.

Figure \ref{fig:someLab}(b) shows the \((11,0)\) SWCT, subjected to an
axial pressure of \(P_{zz}=1 MPa\). Under these conditions, the INT is
predominantly comprised of tetragonal prism blocks, with intermittent
deformed prism blocks.

The length of the INT in Figure \ref{fig:someLab}(a) is smaller than
that in Figure \ref{fig:someLab}(b). The number of prism blocks of each
type is an unreliable indicator, which does not qualitatively describe
the relative proportions of the prismatic ice phases. Although a
volume-based metric has been proposed \cite{Goswami2019}, an approximate
metric based on the normalized number of prism blocks may be used as a
reasonable approximation of relative proportion.

We have observed that the average height of each prism block remains
relatively constant at a value of \(\approx 2.845 \pm 0.07\)
\si{\angstrom}, irrespective of the number of nodes in the basal ring
(\(n\)) and even the applied pressure \(P_{zz}\). Therefore, it is
possible to define a theoretical maximum possible number of \(n\)-gonal
prism blocks, assuming that the entire SWCT height is filled with
\(n\)-gonal prism blocks, each of height \(\approx 2.845\)
\si{\angstrom}. Since the average height of the prism blocks is
independent of \(n\), the theoretical maximum number of prism blocks is
the same for all \(n\). The theoretical maximum number of prism blocks
\(N_{max}\) is thus:

\[N_{max} = \frac{H_{SWCT}}{h_{avg}}\]

where \(H_{SWCT}\) is the height of the SWCT; \(h_{avg}=2.845\) is the
average height of the prism blocks. Both measurements are in
\si{\angstrom}.

The normalized \(height_{n} \%\) for any prismatic ice phase is defined
as follows:

\[Height_{n}\% = \frac{N_{n}}{N_{max}} \times 100\]

where, \(N_{max}\) is the theoretical maximum possible number of
\(n\)-gonal prism blocks; \(N_{n}\) is the actual number of \(n\)-gonal
prism blocks.

Table \ref{prismBlockAmt} summarizes the relative proportions of the
\(n\)-gonal prism blocks for the SWCTs. \(N_{max}\) is the same for
every \(n\), and thus the (11,0) and (13,0) SWCTs have
\(N_{max} = 63.492 \%\) and \(N_{max} = 43.207 \%\), respectively, for
all \(n\). It is evident that the normalized height percentage matches
reasonably well with the occupied volume percentage.

\begin{longtable}[]{@{}lllll@{}}
\caption{Relative proportions of \(n\)-gonal prism blocks
\label{prismBlockAmt}}\tabularnewline
\toprule
\begin{minipage}[b]{0.08\columnwidth}\raggedright
SWCT Type\strut
\end{minipage} & \begin{minipage}[b]{0.14\columnwidth}\raggedright
Actual Number \((N_{n})\)\strut
\end{minipage} & \begin{minipage}[b]{0.16\columnwidth}\raggedright
Maximum Number \((N_{max})\)\strut
\end{minipage} & \begin{minipage}[b]{0.23\columnwidth}\raggedright
Normalized Height \% \((H_{n})\)\strut
\end{minipage} & \begin{minipage}[b]{0.24\columnwidth}\raggedright
Occupied Volume \% \((V_{n})\)\strut
\end{minipage}\tabularnewline
\midrule
\endfirsthead
\toprule
\begin{minipage}[b]{0.08\columnwidth}\raggedright
SWCT Type\strut
\end{minipage} & \begin{minipage}[b]{0.14\columnwidth}\raggedright
Actual Number \((N_{n})\)\strut
\end{minipage} & \begin{minipage}[b]{0.16\columnwidth}\raggedright
Maximum Number \((N_{max})\)\strut
\end{minipage} & \begin{minipage}[b]{0.23\columnwidth}\raggedright
Normalized Height \% \((H_{n})\)\strut
\end{minipage} & \begin{minipage}[b]{0.24\columnwidth}\raggedright
Occupied Volume \% \((V_{n})\)\strut
\end{minipage}\tabularnewline
\midrule
\endhead
\begin{minipage}[t]{0.08\columnwidth}\raggedright
\((11,0)\) SWCT\strut
\end{minipage} & \begin{minipage}[t]{0.14\columnwidth}\raggedright
\(N_4 = 40\)\strut
\end{minipage} & \begin{minipage}[t]{0.16\columnwidth}\raggedright
\(N_{max} = 63.627\)\strut
\end{minipage} & \begin{minipage}[t]{0.23\columnwidth}\raggedright
\(H_{4} = 63.492 \%\)\strut
\end{minipage} & \begin{minipage}[t]{0.24\columnwidth}\raggedright
\(V_{4} = 69.611 \%\)\strut
\end{minipage}\tabularnewline
\begin{minipage}[t]{0.08\columnwidth}\raggedright
\((13,0)\) SWCT\strut
\end{minipage} & \begin{minipage}[t]{0.14\columnwidth}\raggedright
\(N_5 = 4\) \(N_6 = 16\)\strut
\end{minipage} & \begin{minipage}[t]{0.16\columnwidth}\raggedright
\(N_{max} = 43.207\)\strut
\end{minipage} & \begin{minipage}[t]{0.23\columnwidth}\raggedright
\(H_{5} = 9.302 \%\) \(H_{6} = 37.21 \%\)\strut
\end{minipage} & \begin{minipage}[t]{0.24\columnwidth}\raggedright
\(V_{5} = 9.658 \%\) \(V_{6} = 42.982 \%\)\strut
\end{minipage}\tabularnewline
\bottomrule
\end{longtable}

\hypertarget{freezing-of-an-ice-nanotube}{%
\subsubsection{Freezing of an Ice
Nanotube}\label{freezing-of-an-ice-nanotube}}

We track the phase change from the liquid to solid state in water
confined within a \((13,0)\) featureless nanotube, approximating a
zigzag carbon nanotube. The temperature is lowered in steps of
\(10 \ K\). Simulations are run in the \(NP_{zz}T\) ensemble for
\(20 \ ns\) at each temperature, subjected to a constant pressure of
\(0.5 \ MPa\) at each temperature for \(20 \ ns\).

\begin{figure}
\centering
\includegraphics[width=1.0\textwidth]{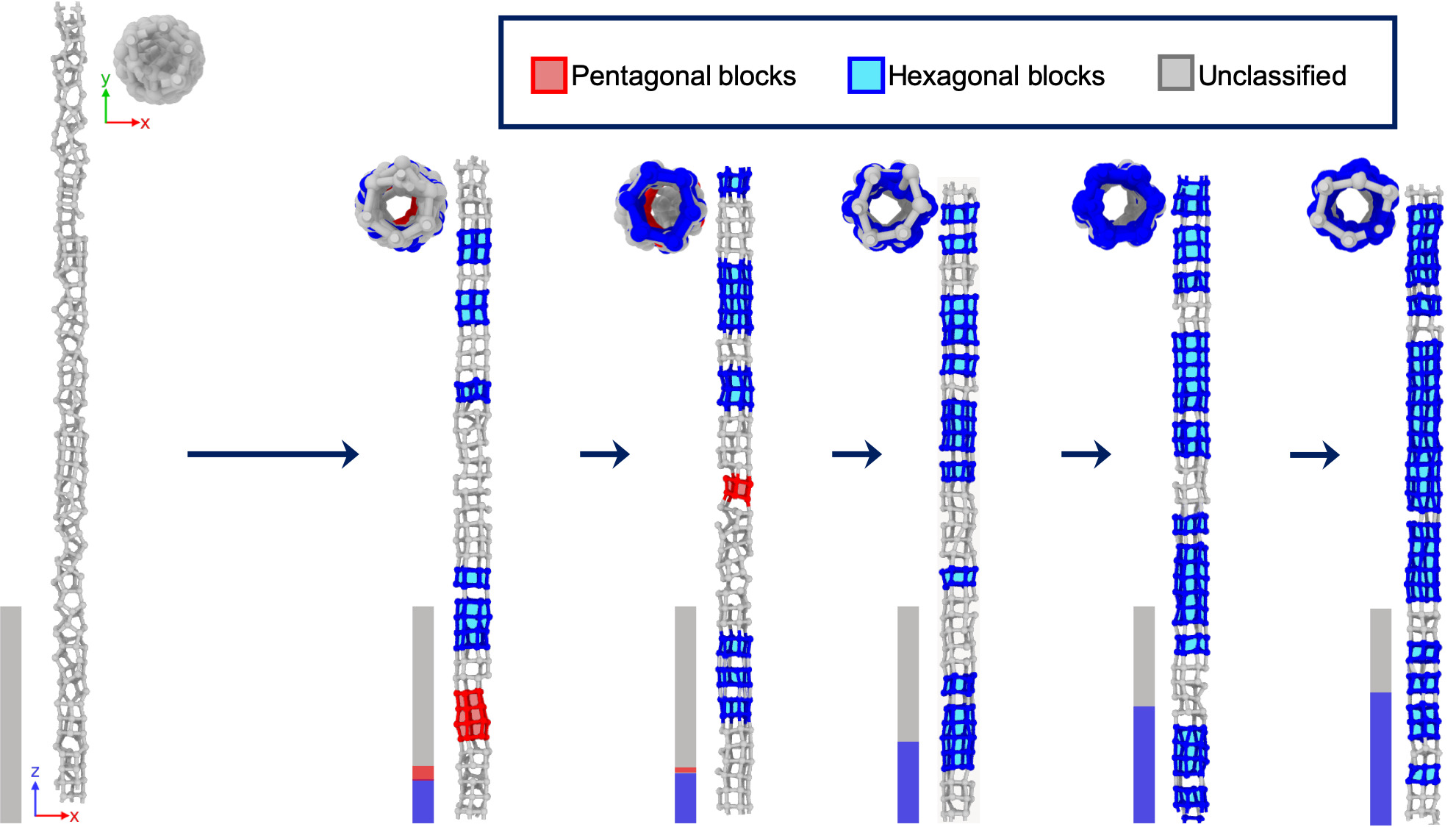}
\caption{Snapshots of an icy nanoribbon at various timesteps, at $280 \
K$ and $P_{zz}=0.5 \ MPa$, in a $(13,0)$ smooth single-walled nanotube. The
stacked bars in the lower left corner of each snapshot denote the height
percentages for the dominant phases: hexagonal prism blocks (blue) and
pentagonal prism blocks (red). Initially, the quasi-one-dimensional water is
entirely in the liquid phase.\label{fig:intEvol}} \end{figure}

We observe an abrupt change in phase from the liquid state to ordered
prismatic ice phase at a temperature of \(280 \ K\). In the previous
section, we showed how the \(height_n \%\) metric matches well with the
occupied volume percentage, which has been proven to be an effective
order parameter for nucleating quasi-one-dimensional systems
\cite{Goswami2019}. Thus, the phase change of an INT can be described by
the \(height_n \%\) order parameter. Figure \ref{fig:intEvol} depicts
the phase change from the liquid phase, at the beginning of the
simulation, to \(\approx 60 \%\) hexagonal prismatic phase after the
\(20 \ ns\) simulation time. The dominant ice phase formed is the
hexagonal prismatic phase, with small proportions of pentagonal ice
(within \(7 \ height_5\%\)).

A movie (entitled \texttt{intEvol.avi} in the ESI) documents this phase
change.

\hypertarget{conclusion}{%
\section{Conclusion}\label{conclusion}}

d-SEAMS is a flexible post-processing analysis tool, capable of
classifying water at both extremes of scale: highly confined systems as
well as bulk water. d-SEAMS is the first scientific software to use
\texttt{nix} as a build-system to circumvent dependency clashes, along
with a \texttt{YAML}-\texttt{Lua} scripting pipeline. The \texttt{Lua}
scripting interface, as well as the \texttt{C++} API, are meant to
provide enough rigor for customization, while being easy to use for the
general scientific community.

Several applications of qualitative analysis have been presented. We
have shown how d-SEAMS is capable of determining the time evolution of
structures, from which the growth mechanism and new physical insights
have been inferred for heterogenous nucleation. A new order parameter
for determining the relative composition of \(n\)-gonal ice in
quasi-one-dimensional nanotubes has been formulated and implemented. The
new order parameter, the \(height \%\), produces results that match well
with previously defined metrics in the literature. We have demonstrated
the versatility of d-SEAMS by performing detailed structural analysis
for homogenous nucleation, fMSI formation and ice nanotube freezing from
the liquid state.

d-SEAMS (https://dseams.info) is a free and open-source molecular
dynamics engine, distributed on GitHub under the GNU General Public
License and documented online (https://docs.dseams.info). We envisage
future development to incorporate more input formats and structural
analysis algorithms. Given the nature of the engine, we expect additions
to the framework which would provide insights into biomolecular systems
as well.

\hypertarget{acknowledgments}{%
\section{Acknowledgments}\label{acknowledgments}}

The computational resources were provided by the HPC facility of the
Computer Center(CC), Indian Institute of Technology Kanpur. R.G.
acknowledges the invaluable support of his lab PI, D. Goswami.

\printbibliography

\end{document}


\hypertarget{design}{%
\section{Design}\label{design}}

The d-SEAMS framework is designed to be accessible to the end-user,
while offering a powerful system of building blocks and generics for
extensions. The engine itself is written in \texttt{C++} and is compiled
to a binary. This binary accepts \texttt{Lua} input scripts to expose
the functionality of the software such that the underlying
data-structures and computations are abstracted away from the user. To
facilitate reproduction of results and to prevent users from accessing
conflicting or unphysical functional manipulations of the input data,
\texttt{YAML} options mask certain functions from being exposed. The
\texttt{YAML} workflows are completely reproducible in \texttt{Lua}
scripts, but provide an easy way to share methodologies and also reduces
the cognitive load of going through the complete API documentation.

\begin{figure}[H]
\centering
\includegraphics[scale=0.5]{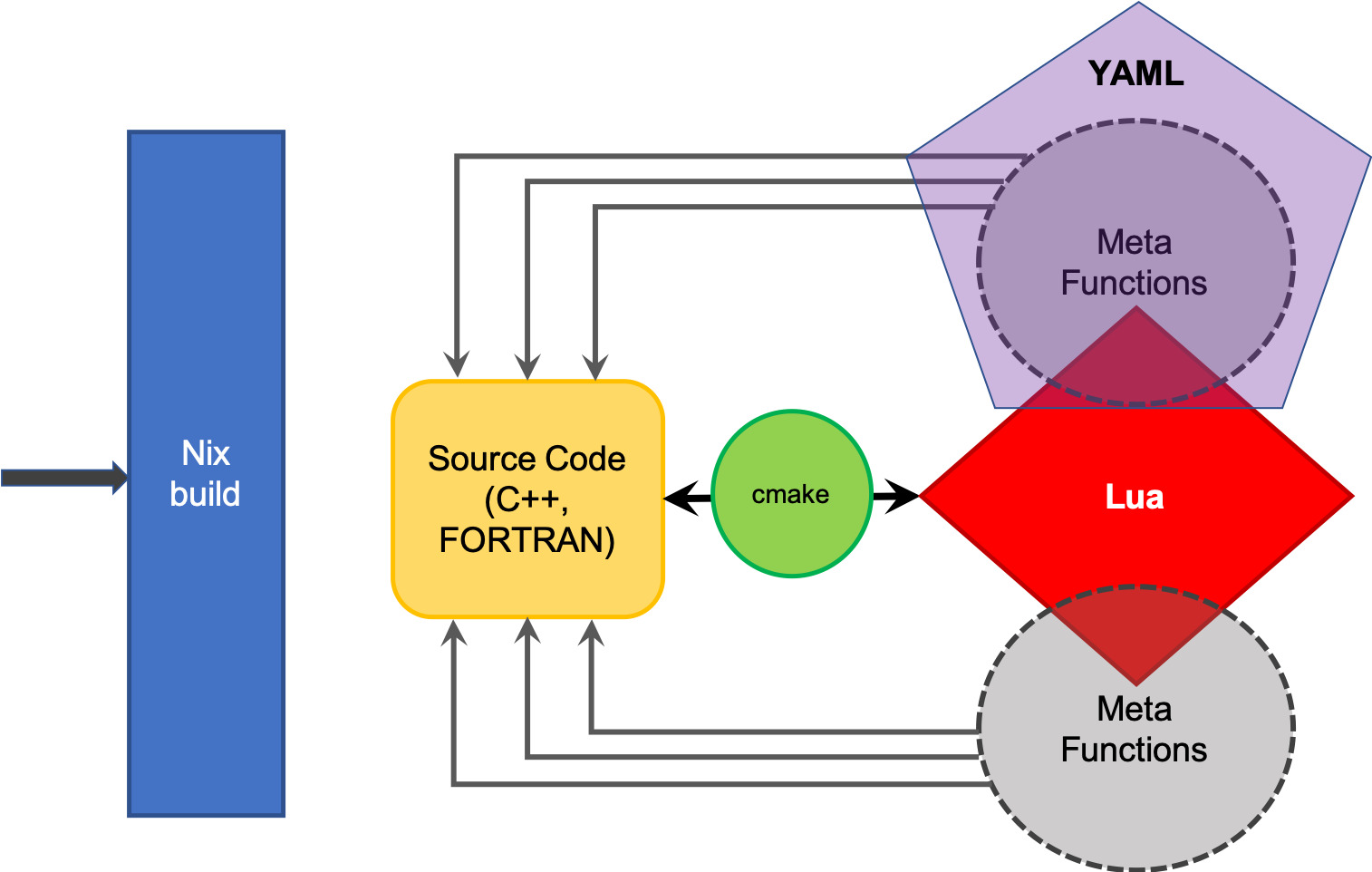}
\caption{Work-flow of d-SEAMS. `Nix` uses cryptographic hashes to ensure
reproducible builds over all systems. `Cmake` compiles and builds the source
code, using the dependencies managed by `nix`. The `Lua` script provides an
interface to the back-end functions. Combinations of these `C++` functions can
be called the 'meta' functions. The `YAML` interface exposes only relevant
back-end and `Lua` functions, corresponding to the user-determined
pre-determined work-flow.}
\label{fig:workFlow}
\end{figure}

Figure \ref{fig:workFlow} is a schematic of the overall design
architecture of d-SEAMS. The \texttt{Lua} scripting interface exposes
\texttt{C}-like functions to create custom work-flows. The \texttt{YAML}
configuration file provides options for pre-determined work-flows. Users
with different requirements and experience can interact with
\texttt{d-SEAMS}. The three main components of the code architecture are
enumerated below:

\begin{itemize}
\item
  \textbf{\texttt{YAML} configuration file:} This contains options for
  pre-determined work-flows. For example, a user who enters the option
  for the confined quasi-one-dimensional ice determination will only be
  exposed to the relevant functions for INT prism determination.
  Multiple workflows can be selected at the same time.
\item
  \textbf{\texttt{Lua} Interface:} C++ functions are registered as
  \texttt{Lua} functions, which are called from a \texttt{Lua} script.
  \texttt{Lua} is a \texttt{C}-like scripting language, enabling users
  to call the \texttt{Lua} functions without needing to learn a
  software-specific scripting convention. The advantage of using
  \texttt{Lua} over directly calling \texttt{C++} functions is that the
  users need not be concerned with pointers and clean-up of the
  \texttt{C++} structures. The \texttt{Lua} language also has a rich set
  of cross platform extensions for file handling, and is also supported
  by major editors for syntax highlighting.
\item
  \textbf{\texttt{C++} Back-end:} The back-end is written in modern
  \texttt{C++}, employing common data structures, used uniformly
  throughout the code. Users can easily extend and write their own
  \texttt{C++} header files, and the documentation covers manipulating
  the build system to accept both user-defined and external headers.
  Registering custom \texttt{C++} functions as \texttt{Lua} functions,
  to be subsequently called in \texttt{Lua} scripts, is also documented.
  \texttt{GDB} \cite{stallman2002debugging} can be used for code
  debugging, since the back-end is in \texttt{C++}.
\end{itemize}

From a user perspective, we have designed the Lua functions to mimic the
mindset of a computational chemist, without burdening them with the
software implementation. We have also ensured reproducibility, both as
an aid to the science \cite{mesirovAccessibleReproducibleResearch2010}
intended and also to allow for bugs to be dealt with more efficiently.
This reproducibility is ensured during build, compile, and linking
stages, by leveraging the functional, immutable binaries produced by nix
\cite{dolstraNixSafePolicyFree2004}. The dependencies are handled
reproducibly, though for ease of extension by the wider community, most
of the build system is in CMake. We use nix to ensure that the
dependencies of the binary are fully reproducible, as a consequence of
traversing the build graph defined by the nix-derivation. The binary
itself has a server-client architecture, to ensure that the user can
transparently interact with the code without needing a background in
functional programming. Since the backend server is written entirely in
modern C++, the GDB debugger is usable throughout. The server-client
nature of the system, though currently a bottleneck in terms of
parallelism, allows for a single compiled binary to be used for the
execution of multiple different Lua input scripts, with each script
spawning a separate process.

\hypertarget{nix}{%
\subsection{Nix}\label{nix}}

The nix derivation provides a deterministic package-level lock on
all dependencies and is written in \texttt{nix}, a lazy, dynamically
typed, purely functional language
\cite{dolstraNixOSPurelyFunctional2010}. This choice of distribution
also allows the user to extend the system reproducibly, ensuring that
changes can be quickly merged in-to the upstream repository. 
More details of the `nix`-build system are described in the main text of 
the manuscript.

\hypertarget{lua}{%
\subsection{Lua}\label{lua}}

Existing molecular dynamics packages suffer from not having design
parameters built-in, and with time, this has led to unique and
non-standard syntax being used, as seen in the input scripts of LAMMPS
\cite{plimptonFastParallelAlgorithms} and GROMACS
\cite{abrahamGROMACSHighPerformance2015}, amongst others. Popular text
editors do not offer syntax highlighting for these custom non-standard
and software-specific syntaxes. For such software-specific syntaxes, the
code is unusable without learning from the documentation. An alternative
to crafting a new input script syntax for each software, is to use
\texttt{python} for scripting \cite{mcgibbonMDTrajModernOpen2015}.
However, the version dependence of each internal segment can become
intractable without continuous development, and as a result, these spawn
multiple language-specific errors, and they work best only on the Linux
distribution on which the creators have worked . Furthermore, we assert
that the proliferation of \texttt{python} scripts and the odd-ease at
which they may be, in theory mixed and matched, in practice causes many
clashes, for example, EsPreSSo \cite{weikESPResSoExtensibleSoftware2019}
and Quantum EsPreSSo \cite{giannozziQUANTUMESPRESSOModular2009} (and
more generally, Scipy \cite{oliphantPythonScientificComputing2007} and
Numpy \cite{waltNumPyArrayStructure2011}) have several function names in
common which cause difficult to debug when used together in an input
script. Also, \texttt{python} has less support for debugging, and the
language server support is lacking, making complex \texttt{python} code
difficult to debug. This is partially due to the design of
\texttt{python} itself. The lack of static typing has the effect of
making the compiler oblivious to bugs which are caught easily in
statically typed languages like \texttt{FORTRAN} and \texttt{C++}.

To address these concerns we have opted to use \texttt{Lua} as the
scripting interface, which has \texttt{C}-like functions. It is widely
supported in terms of syntax highlighting, and is easy to interface with
\texttt{C++} code. Furthermore, the error handling is such that it is
amenable for arbitrarily complex GDB debugger workflows
\cite{stallmanDebuggingGDB}, and the rich standard library of
\texttt{Lua}, along with user extensions, have no clashes. \texttt{Lua}
is also user friendly due to its C-like syntax. The rich table and
object handling makes writing out image data easy. Furthermore, though
\texttt{Lua} was released two years after python was first introduced
(1993 and 1991, respectively) unlike \texttt{python}, which is still
transitioning from the major API change of two to three, the
\texttt{Lua} API is stable, mostly because it has been designed to be an
embedded language and not a general-purpose language like python.
\texttt{Lua} has been the darling of the gaming development community,
and has proven its worth in many related domains such as image handling.
Apart from the user-friendly helper functions, our design has the
\texttt{Lua} interface, which offers every core function to the user.
This permits arbitarily complicated workflows to be used without
re-compiliation, which is a boon for HPC cluster usage. We recommend
strongly in the docs, that foreign code, once interfaced to the C++
engine, should be bound in \texttt{Lua} for the end-users as well.

\begin{figure}[H]
\centering
\includegraphics[scale=0.5]{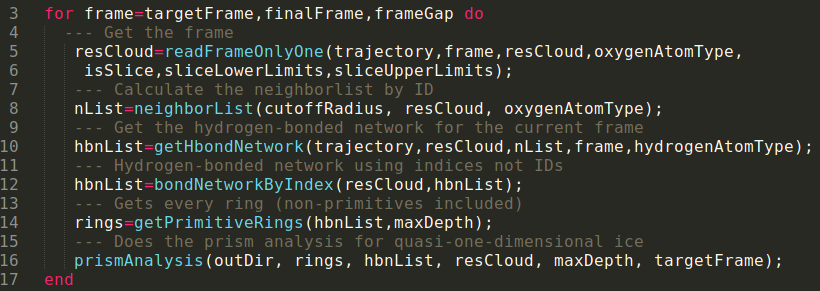}
\caption{\small The `lua` input script, where the user is able to call any of
the functions not voided by the options in the `YAML` file.\label{fig:sampleLua} \normalsize}
\end{figure}

Figure \ref{fig:sampleLua} shows a typical \texttt{Lua} input script,
which calls functions exposed by the current \texttt{YAML} file
work-flow.

\hypertarget{yaml}{%
\subsection{YAML}\label{yaml}}

\begin{figure}[H]
\centering
\includegraphics[width=0.5\textwidth]{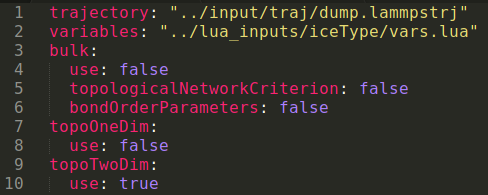}
\caption{\small The `YAML` file, where boolean values are set to restrict
functions exposed to the `lua` scripting engine.\label{fig:sampleYML} \normalsize}
\end{figure}

To improve usability and reduce the time required reading the API
documentation, we have split the usage into a unique YAML-Lua design.
The \texttt{Lua} interface is for power users, however, to reduce
mistakes, options set in the YAML files will deactivate certain
functions, in order to prevent incorrect manipulations of the internal
data-structures. The \texttt{YAML} interface diverts the flow of
functionality and code to different paths, and thus different
algorithms. This also prevents name-clashes of similar functions for
mutually exclusive work-flows. For example, an input system can either
be a bulk system, a quasi-one-dimensional system or a
quasi-two-dimensional system. The \texttt{YAML} file offers truthy
options, an example of which is shown in Figure \ref{fig:sampleYML}, and
subsequently masks functions not applicable for the given system type.

\bibliography{refs.bib}